\documentclass[graphics,floatfix, nofootinbib,tightenlines,nobibnotes, 12pt]{revtex4-1}
\usepackage{graphicx}
\usepackage[english]{babel}
\usepackage[utf8]{inputenc}
\usepackage{amsmath,amssymb}
\usepackage{amsfonts}
\usepackage{multirow}
\usepackage{pstricks}
\usepackage{float}
\usepackage{subfigure}
\usepackage{color}
\usepackage{epsfig}
\usepackage{slashed}
\usepackage[colorlinks=true, linkcolor=blue, citecolor=blue, urlcolor=blue]{hyperref}

\begin{document}
\title{Production of doubly heavy baryons via Higgs boson decays}

\author{Juan-Juan Niu$^{1}$}
\email{niujj@cqu.edu.cn}
\author{Lei Guo$^{1}$}
\email{guoleicqu@cqu.edu.cn, correponding author}
\author{Hong-Hao Ma$^{2}$}
\email{mahonghao.br@gmail.com}
\author{Xing-Gang Wu$^{1}$}
\email{wuxg@cqu.edu.cn}

\address{$^{1}$ Department of Physics, Chongqing University, Chongqing 401331, People's Republic of China}
\address{$^{2}$ Faculdade de Engenharia de Guaratinguet\'a, Universidade Estadual Paulista, Guaratinguet\'a, SP, 12516-410, Brazil}

\date{\today}

\begin{abstract}
We systematically analyzed the production of semi-inclusive doubly heavy baryons ($\Xi_{cc}$, $\Xi_{bc}$ and $\Xi_{bb}$) for the process $H^0 \rightarrow \Xi_{QQ'}+ \bar {Q'} + \bar {Q}$ through four main Higgs decay channels within the framework of non-relativistic QCD. The contributions from the intermediate diquark states, $\langle cc\rangle[^{1}S_{0}]_{\mathbf{6}}$, $\langle cc\rangle[^{3}S_{1}]_{\mathbf{\bar 3}}$, $\langle bc\rangle[^{3}S_{1}]_{\mathbf{\bar 3}/ \mathbf{6}}$, $\langle bc\rangle[^{1}S_{0}]_{\mathbf{\bar 3}/ \mathbf{6}}$, $\langle bb\rangle[^{1}S_{0}]_{\mathbf{6}}$ and $\langle bb\rangle[^{3}S_{1}]_{\mathbf{\bar 3}}$, have been taken into consideration. The differential distributions and three main sources of the theoretical uncertainties have been discussed. At the High Luminosity Large Hadron Collider, there will be about 0.43$\times10^4$ events of $\Xi_{cc}$, 6.32$\times10^4$ events of $\Xi_{bc}$ and 0.28$\times10^4$ events of $\Xi_{bb}$ produced per year. There are fewer events produced at the Circular Electron Positron Collider and the International Linear Collider, about $0.26\times 10^{2}$ events of $\Xi_{cc}$, $3.83\times 10^{2}$ events of $\Xi_{bc}$ and $0.17\times 10^{2}$ events of $\Xi_{bb}$ in operation.


\end{abstract}

\maketitle

\section{Introduction}

The Higgs boson, as the last found fundamental particle in the standard model (SM), is of great interest to the experimenter and theorist of particle physics. Some future colliders that can be called ``Higgs factories'' would generate large amounts of Higgs particles. The High Luminosity LHC (HL-LHC) running at center-of-mass collision energy $\sqrt{s}=14$ TeV with the integrated luminosity of $3~ab^{-1}$ would produce about $1.65\times10^{8}$ events of Higgs boson per year \cite{LHCHIGGS}; the Circular Electron Positron Collider (CEPC) would generate more than one million Higgs particles at the center-of-mass energy of 240 GeV with the integrated luminosity of $0.8~ab^{-1}$ in 7 years~\cite{CEPCStudyGroup:2018rmc}; and the International Linear Collider (ILC) would generate almost the same magnitude of Higgs bosons as the CEPC, about $10^{5}$-$10^{6}$ at each energy stage~\cite{Simon:2012ik}. Therefore, the decay of Higgs boson will be a good platform for studying the indirect production mechanism of doubly heavy hadrons. Many pioneering investigations of the production of doubly heavy meson through Higgs boson's decay have been done not only by experimental groups but also by theorist, i.e., the production of $B_c$, $J/\psi$ and $\Upsilon$~\cite{Aad:2015sda,Achasov:1991ms,Bodwin:2014bpa,Koenig:2015pha,Qiao:1998kv,Jiang:2015pah,Liao:2018nab}.
The analysis of Higgs boson decays also provides a platform for seeking the undetected doubly heavy baryons.
The doubly heavy baryon contains two heavy quarks and a light quark as valence quarks. For convenience, $\Xi_{QQ^{\prime}}$ is used to stand for the doubly heavy baryons $\Xi_{QQ^{\prime}q_{l}}$ in this paper, where $Q$ and $Q^{\prime}$ represent the heavy quarks ($c$ or $b$ quark) and $q_{l}$ denotes the light quark ($u$, $d$ or $s$ quark). A careful study of the production of doubly heavy baryons $\Xi_{QQ'}$ through Higgs boson decays shall be helpful for confirming whether enough baryon events could be produced and supporting forward guidance on the experiment research.

Attributed to the first observation of the doubly charm baryon $\Xi_{cc}^{++}$~\cite{Aaij:2017ueg} by the LHCb collaboration in 2017, the quark model has proved to be a great success~\cite{GellMann:1964nj,Zweig:1981pd,Zweig:1964jf,DeRujula:1975qlm}. However, there is no explicit evidence of the other doubly heavy baryons $\Xi_{bc}$ and $\Xi_{bb}$ so far. To study all possible production mechanisms of doubly heavy baryons shall be helpful for better understanding their properties and shall be a verification of the quark model and non-relativistic Quantum Chromodynamics (NRQCD) \cite{Bodwin:1994jh,Petrelli:1997ge}. There were some analyses of the direct/indirect production of doubly heavy baryons through $e^+~e^-$ colliders~\cite{Kiselev:1994pu,Ma:2003zk,Zheng:2015ixa,Jiang:2012jt}, hadronic production~\cite{Berezhnoy:1995fy,Doncheski:1995ye,Baranov:1995rc,Berezhnoy:1998aa,Ma:2003zk,Chang:2006eu,Chang:2007pp,Chang:2009va,Zhang:2011hi,Wang:2012vj,Chen:2014hqa}, gamma-gamma production~\cite{Baranov:1995rc,Li:2007vy}, photoproduction~\cite{Baranov:1995rc,Chen:2014frw,Huan-Yu:2017emk}, heavy ion collisions~\cite{Yao:2018zze,Chen:2018koh}, top quark decays~\cite{topdecay}, etc.

In this paper, we shall discuss the production of doubly heavy baryons $\Xi_{QQ'}$ through indirectly Higgs boson decays at the HL-LHC and CEPC/ILC. As is well-known, the dominant decay channel of Higgs boson is $H^0 \rightarrow b\bar{b}$ and the branching ratio is about 58$\%$~\cite{Patrignani:2016xqp,Niu:2018otv}. For completeness, four main Higgs decay channels, $H^0\rightarrow b\bar{b}$, $c\bar{c}$, $Z^0 Z^0$, $gg$, would be taken into consideration. Due to the Yukawa coupling and the perturbative order, the decay channel $H^0 \rightarrow b\bar{b}$ ($H^0 \rightarrow c\bar{c}$) plays an essential role in the production of $\Xi_{bc}$ and $\Xi_{bb}$ ($\Xi_{cc}$), but the contributions from the $H^0\rightarrow Z^0 Z^0/ gg$ channel cannot be neglected.

Within the framework of NRQCD, the production of doubly heavy baryons $\Xi_{QQ^{\prime}}$ can be factorized into the convolution of the perturbative short-distance coefficient and the non-perturbative long-distance matrix elements.
In the amplitude, the gluon is hard enough to produce such a heavy quark-antiquark pair, hence the hard process is perturbatively calculable.
The long-distance matrix elements are used to describe the transition probability of the produced diquark state $\langle QQ^{\prime} \rangle[n]$ binding
into doubly heavy baryons $\Xi_{QQ^{\prime}}$, where $[n]$ stands for the spin and color quantum number for the intermediate diquark state. The spin quantum number of the intermediate diquark state $\langle QQ^{\prime}\rangle[n]$ can be $[^3S_1]$ or $[^1S_0]$, and the color quantum number is the color-antitriplet $\bar 3$ or the color-sextuplet 6 for the decomposition of $SU_C(3)$ color group $3\bigotimes3=\bar{3} \bigoplus6$.
All of these intermediate states would be taken into consideration for a sound estimation. Assuming the potential of the binding color-antitriplet $\langle QQ'\rangle[n]$ state is hydrogen-like, the transition probability $h_{\bar 3}$ can be approximatively related to the Schr\"{o}dinger wave function at the origin $|\Psi_{QQ'}(0)|$ for the $S$-wave states, where $|\Psi_{QQ'}(0)|$ can be obtained by fitting the experimental data or some non-perturbative methods like QCD sum rules~\cite{Kiselev:1999sc}, lattice QCD~\cite{Bodwin:1996tg} or the potential model~\cite{Bagan:1994dy}. As for the transition probability of the color-sextuplet diquark state $h_{6}$, there is a relatively larger uncertainty, and we would make a detailed discussion about it.

The remaining parts of the paper are arranged as follows: in Sect. 2, the detailed calculation technology, such as the factorization and the color factors, is presented. The numerical results associated with the theoretical uncertainties are given in Sect. 3. And Sect. 4 gives a summary and some conclusions.

\section{Calculation technology}

\begin{figure}[htb]
  \centering
  \subfigure[]{
    \includegraphics[scale=0.27]{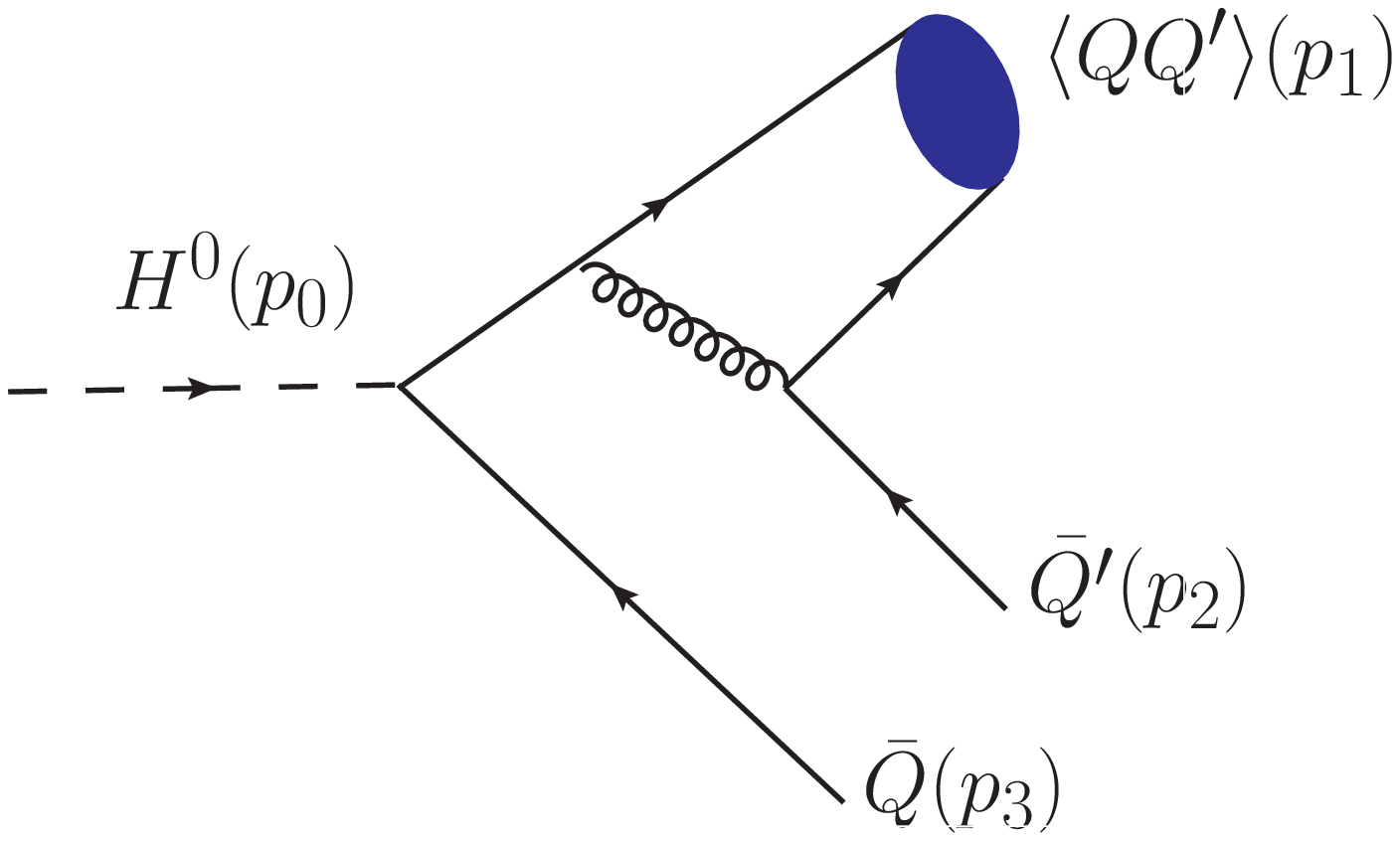}}
  \hspace{0.00in}
  \subfigure[]{
    \includegraphics[scale=0.27]{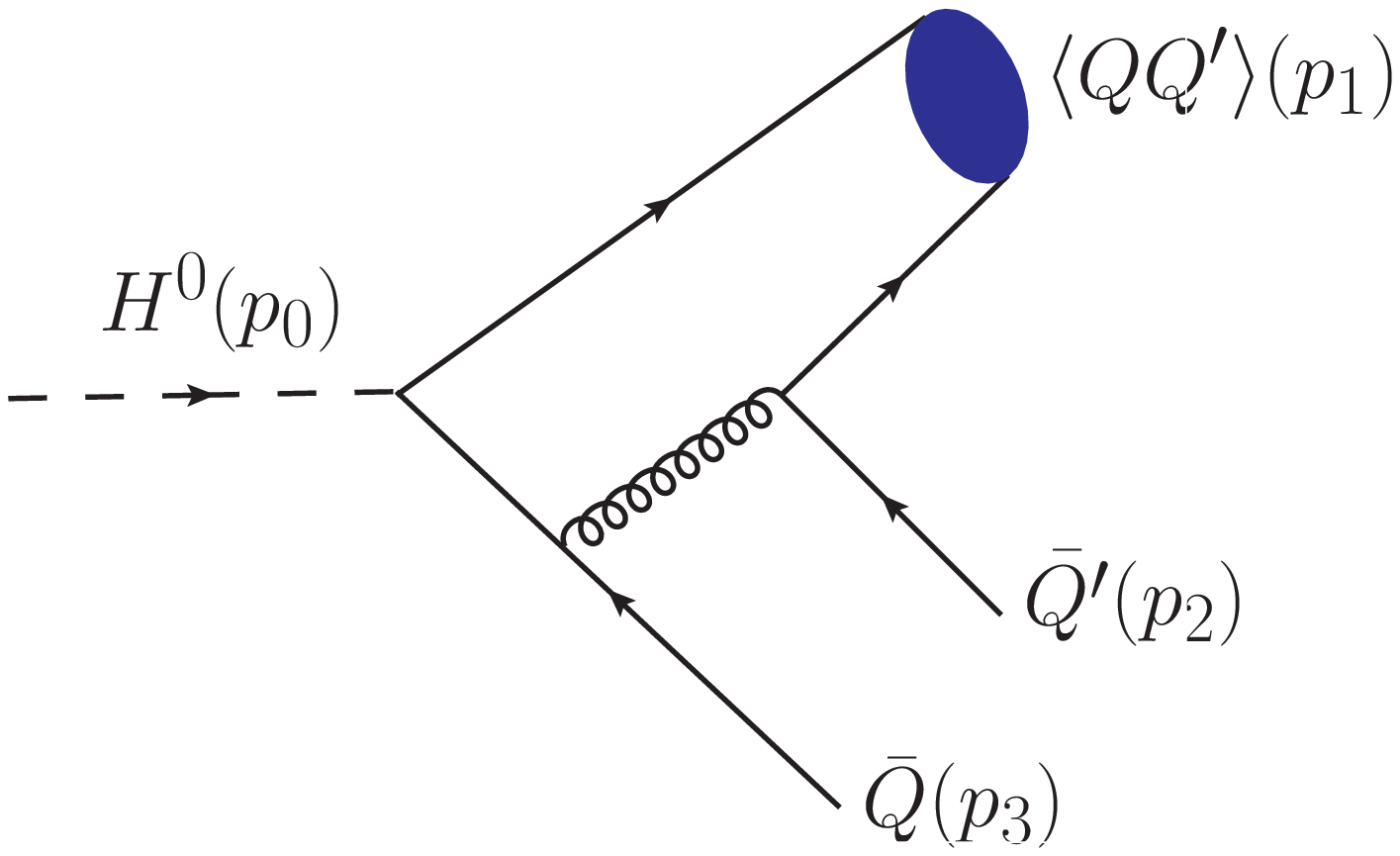}}
  \hspace{0.00in}
  \subfigure[]{
    \includegraphics[scale=0.33]{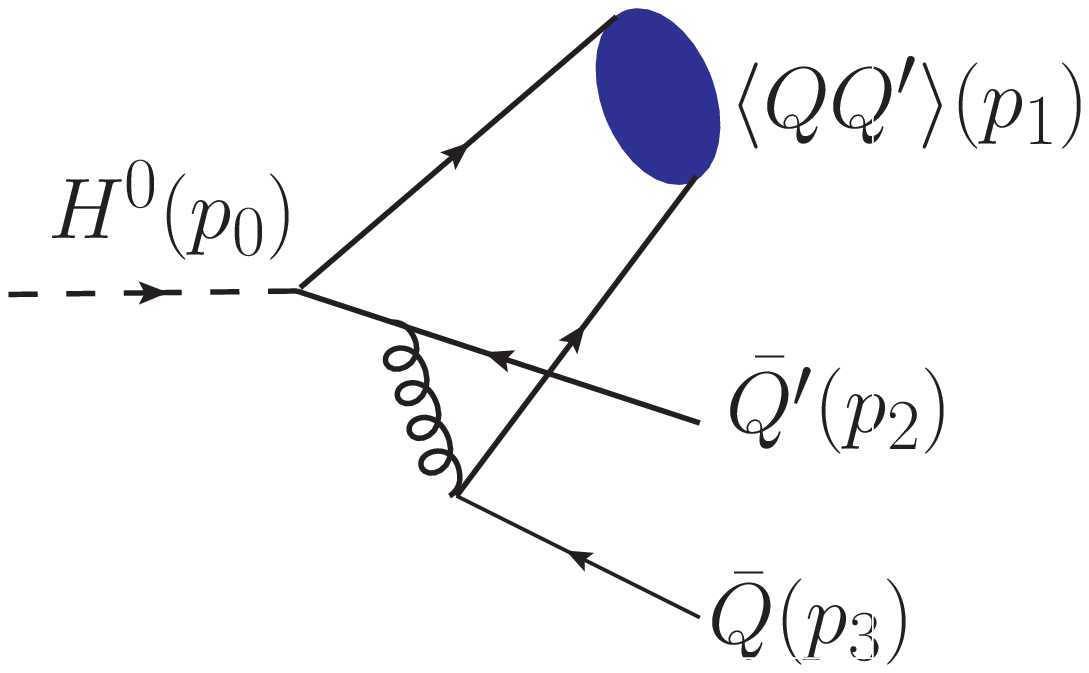}}
  \hspace{0.00in}
  \subfigure[]{
    \includegraphics[scale=0.31]{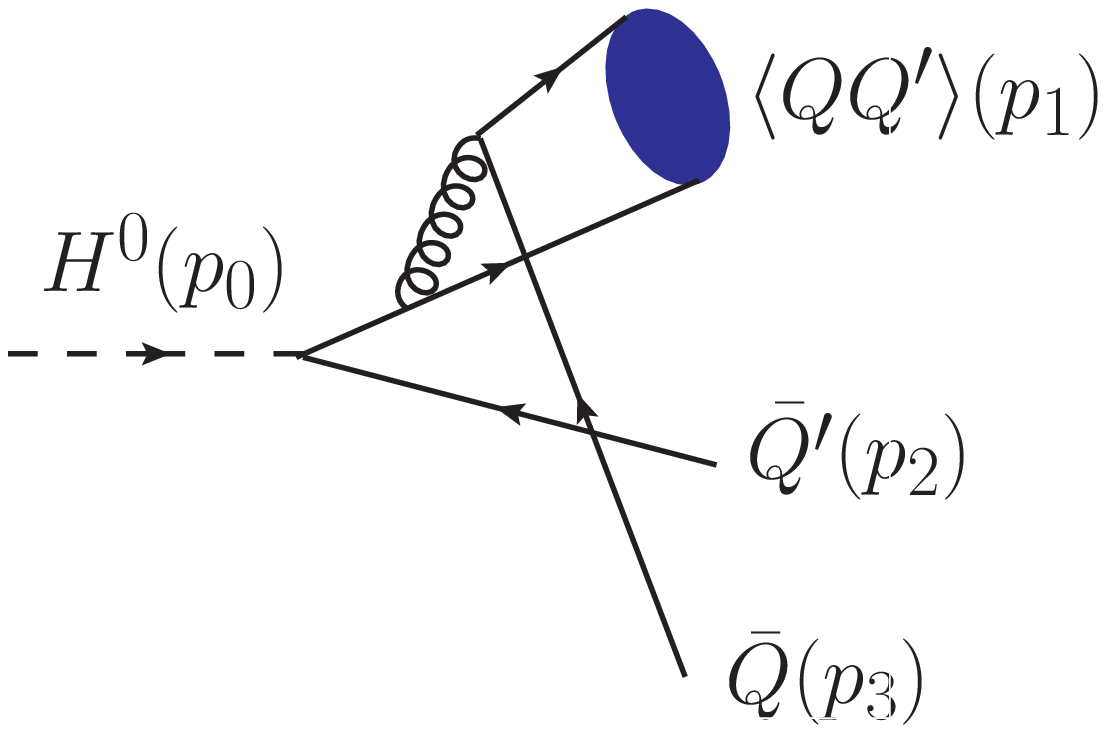}}
     \hspace{0.00in}
   \subfigure[]{
    \includegraphics[scale=0.3]{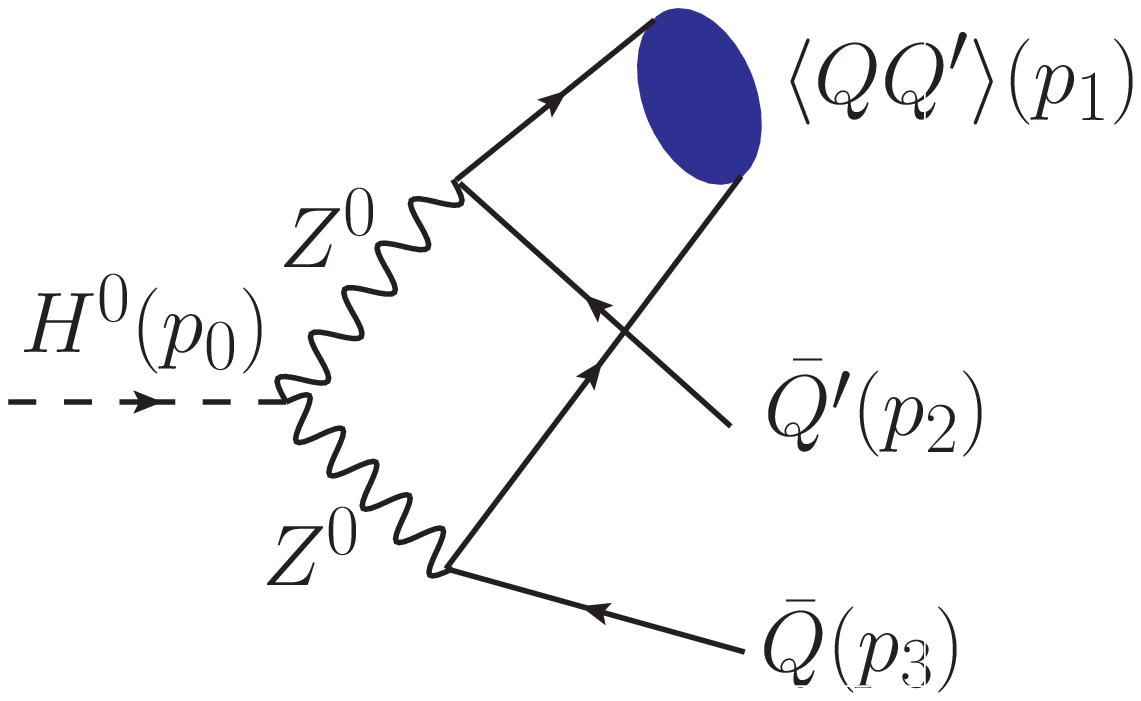}}
    \hspace{0.00in}
  \subfigure[]{
    \includegraphics[scale=0.3]{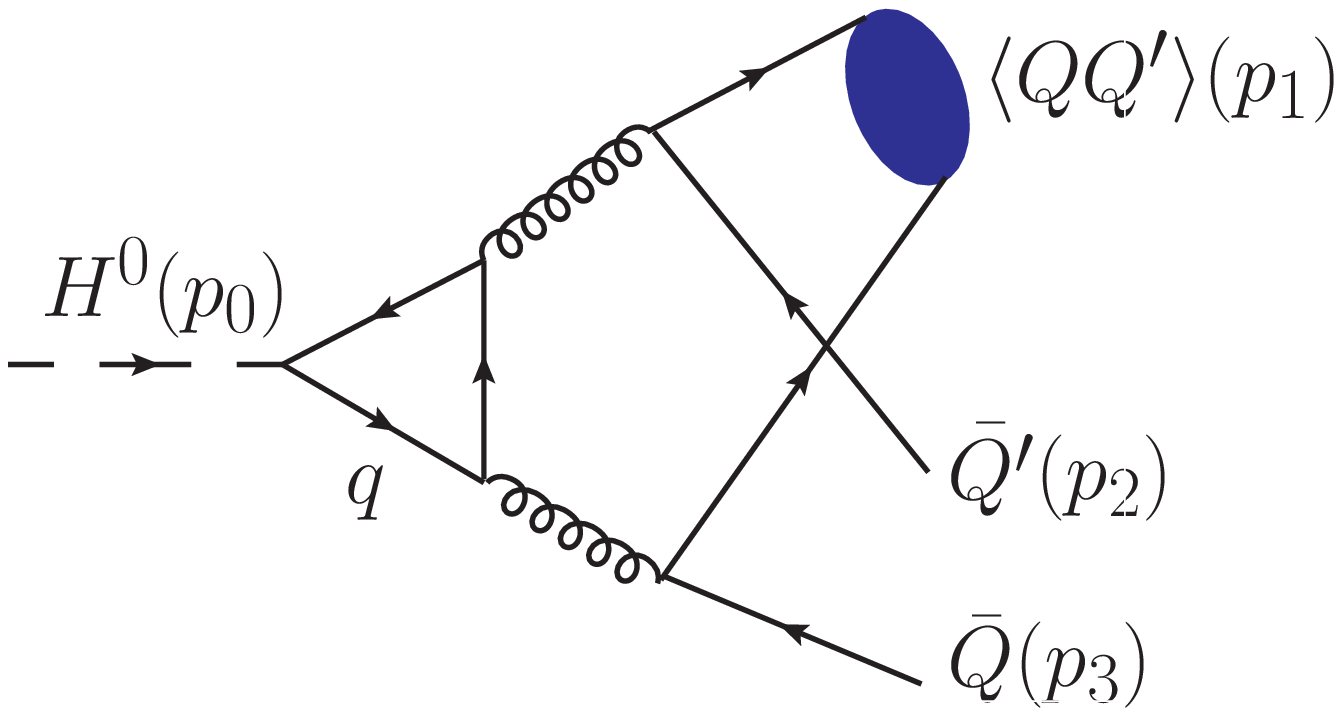}}
  \hspace{0.00in}
  \subfigure[]{
    \includegraphics[scale=0.3]{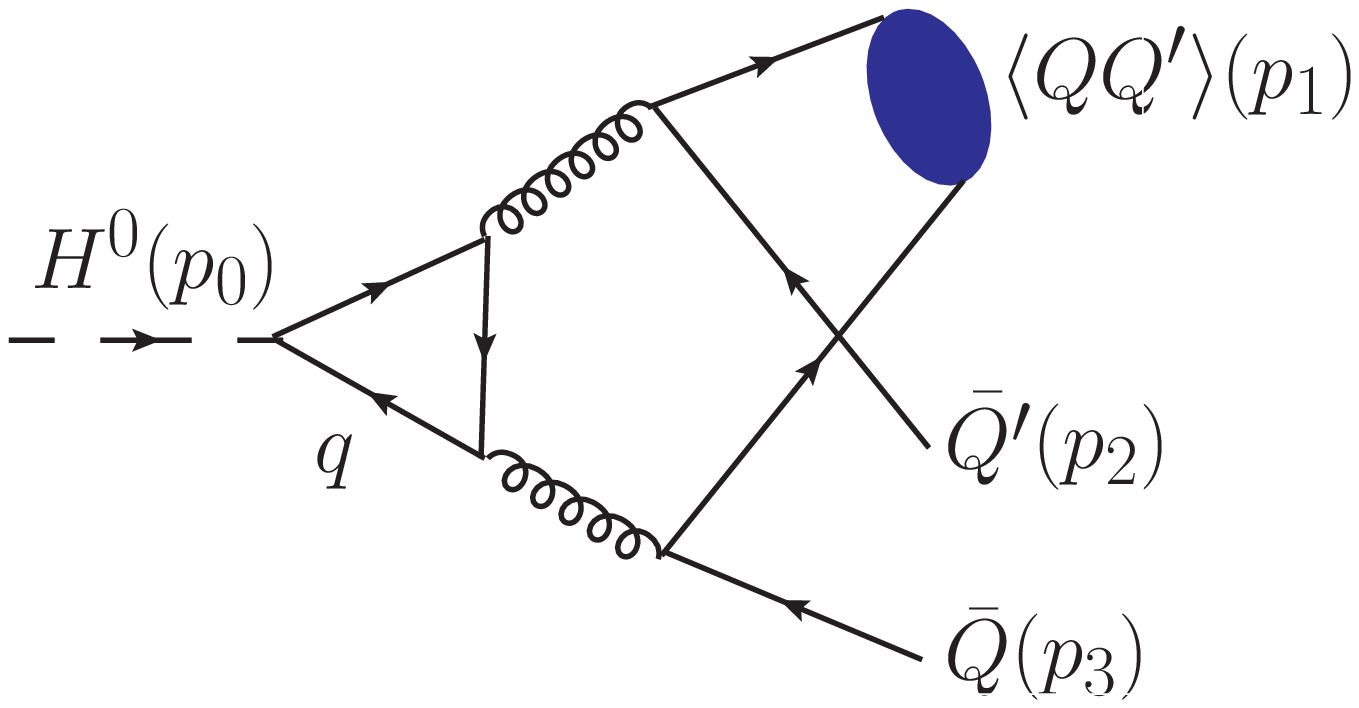}}
  \caption{Typical Feynman diagrams for the process $H^0 (p_0) \rightarrow Q\bar{Q}/Q^{\prime}\bar{Q^{\prime}}/Z^{0}Z^{0} / gg \rightarrow \Xi_{QQ^{\prime}} (p_1)+ \bar {Q^{\prime}} (p_2) + \bar{Q} (p_3)$, where $Q$ and $Q^{\prime}$ denote as the heavy $c$ or $b$ quark.}
  \label{diagram1}
\end{figure}
Typical Feynman diagrams for the process $H^0 (p_0)\rightarrow \Xi_{QQ^{\prime}}(p_1)+ \bar{Q^{\prime}}(p_2) + \bar{Q}(p_3)$ through four main Higgs decay channels, $H^0 \rightarrow b\bar{b}/c\bar{c}/Z^{0}Z^{0} / gg$, are presented in Fig.~\ref{diagram1}, where $Q$ and $Q^{\prime}$ denote the heavy $c$ or $b$ quark for the production of $\Xi_{cc}$, $\Xi_{bc}$ and $\Xi_{bb}$ accordingly. Within the framework of NRQCD~\cite{Bodwin:1994jh,Petrelli:1997ge}, the decay width for the production of $\Xi_{QQ'}$ can be factorized as the following form:
\begin{eqnarray}
\Gamma &&(H^0(p_0) \rightarrow \Xi_{QQ^{\prime}}(p_1)+ \bar {Q^{\prime}}(p_2) + \bar {Q}(p_3)) \nonumber\\
&&=\sum_{n} \hat{\Gamma}(H^0(p_0) \rightarrow \langle QQ^{\prime}\rangle[n](p_1) + \bar {Q^{\prime}}(p_2) + \bar {Q}(p_3)) \langle\mathcal O^{H}[n]\rangle,
\end{eqnarray}
where the non-perturbative long-distance matrix element $\langle\mathcal O^{H}[n]\rangle$ is proportional to the transition probability from the perturbative quark pair $\langle QQ^{\prime}\rangle[n]$ to the heavy baryons $\Xi_{QQ^{\prime}}$. According to NRQCD, the intermediate diquark state $\langle QQ^{\prime}\rangle[n]$ can be expanded to a series of Fock states with different spin and color quantum numbers $[n]$, which are accounted by the velocity scaling rule. Due to the symmetry of identical particles in the diquark state, the intermediate diquark state can be either $[^3S_1]_{\bar 3}$ or $[^1S_0]_{6}$ for the production of $\Xi_{cc}$ and $\Xi_{bb}$ baryons. Meanwhile, for the production of baryon $\Xi_{bc}$ there are four color and spin states such as $\langle bc\rangle[^3S_1]_{\bar{3}}$, $\langle bc\rangle[^1S_0]_{\bar{3}}$, $\langle bc\rangle[^3S_1]_{6}$ and $\langle bc\rangle[^1S_0]_{6}$. All of these Fock states would be taken into consideration for a comprehensive understanding. We shall use $h_{\bar{3}}$ and $h_6$ to describe the transition probability of the color-antitriplet diquark state and the color-sextuplet diquark state, respectively. In addition, the transition probability $h_{\bar 3}$ can be approximatively related to the Schr\"{o}dinger wave function at the origin $|\Psi_{QQ'}(0)|$ for the $S$-wave states, while there is a relatively larger uncertainty for the transition probability $h_{6}$, which has been analyzed detailedly in Ref.~\cite{topdecay}. For convenience, we set $h_{6} \simeq h_{\bar 3}=|\Psi_{QQ^{\prime}}(0)|^2$ \cite{Bagan:1994dy,Petrelli:1997ge} as an approximate estimate.

The decay width $ \hat{\Gamma}(H^0 \rightarrow \langle QQ^{\prime}\rangle[n] + \bar {Q^{\prime}} + \bar {Q})$ represents the perturbative short-distance coefficients which can be written as
\begin{eqnarray}
\hat{\Gamma}(H^0 \rightarrow \langle QQ^{\prime}\rangle[n] + \bar {Q^{\prime}} + \bar {Q})= \int \frac{1}{2m_H} \sum |\mathcal{M}[n]|^2 d\Phi_3,
\label{width}
\end{eqnarray}
where $m_H$ is the mass of the Higgs boson, $\mathcal{M}[n]$ is the hard amplitude, and $\sum$ means to sum over the spin and color of the final-state particles. The three-body phase space $d\Phi_3$ can be expressed as
\begin{eqnarray}
d\Phi_3=(2\pi)^4 \delta^4(p_0-\sum_{f=1}^{3} p_f) \prod_{f=1}^{3} \frac{d^3 p_f}{(2\pi)^3 2 p_{f}^{0}}.
\end{eqnarray}
After performing the integration over the phase space $d\Phi_3$, Eq.~(\ref{width}) can be rewritten as
\begin{eqnarray}
d \hat{\Gamma} = \frac{1}{256 \pi^{3} m_{H}^{3}} \sum |\mathcal{M}[n]|^2 ds_{12} ds_{23},
\end{eqnarray}
where the definitions of the invariant mass are $s_{ij}=(p_{i}+p_{j})^{2}$, ($i,j=1,2,3$). Therefore, not only the total decay width but also the corresponding differential distributions can be derived, which are helpful for experimental measurements.

\subsection{Amplitude}

We made a relatively complete analysis for the production of doubly heavy baryons $\Xi_{QQ'}$ through four main Higgs decay channels, $H^0 \rightarrow b \bar{b},~c \bar{c}, ~Z^{0} Z^{0}$, and $gg$. Subgraphs (a)-(d) in Fig.~\ref{diagram1} are specifically represented the channels $H^0 \rightarrow b \bar{b}/c \bar{c}\rightarrow \Xi_{QQ^{\prime}}+ \bar{Q^{\prime}} + \bar{Q}$, while subgraph (e) is for $H^0 \rightarrow Z^{0} Z^{0}\rightarrow \Xi_{QQ^{\prime}}+ \bar{Q^{\prime}} + \bar{Q}$ and subgraphs (f)-(g) represent the decay channel $H^0 \rightarrow gg \rightarrow \Xi_{QQ^{\prime}}+ \bar{Q^{\prime}} + \bar{Q}$. In subgraphs (f) and (g), $q$ stands for $t,~b$ or $c$ quark. According to the Yukawa coupling, the top quark in the triangle loop can make the largest contribution to the decay width through $H^0 \rightarrow gg$. After the action of the charge parity $C=-i\gamma^2\gamma^5$, the hard amplitude $\mathcal M[n]$ for the production of the intermediate diquark state can be related to the familiar meson production, which has been proved in Refs.~\cite{Jiang:2012jt,Zheng:2015ixa} in detail. In other words, we could obtain the hard amplitude $\mathcal M[n]$ of the process $H^0(p_0) \rightarrow \langle QQ^{\prime}\rangle[n](p_1) + \bar Q^{\prime}(p_2) + \bar{Q}(p_3)$ from the process $H^0(p_0) \rightarrow (Q\bar {Q^{\prime}})[n](p_1) + Q^{\prime}(p_2) + \bar{Q}(p_3)$ with an additional factor $(-1)^{m+1}$, where $m$ stands for the number of vector vertices in the $Q'$ fermion line which need to be reversed and here $m$ = 1 for these four decay channels. It is worth mentioning that there are vector and axial vector contributions in the channel $H^0 \rightarrow Z^0Z^0$, and $m$ = 0 for the axial vector contribution. Additionally, the $Z^0$ propagators should be considered as Breit-Wigner propagators to avoid the resonance.

According to Fig.~\ref{diagram1}, the hard amplitude $\mathcal M_l[n]~(l=a,\ldots,g)$ can be written as
\begin{widetext}
\begin{eqnarray}
\mathcal{M}_{a}[n]&=&\mathcal{C} \bar{u}_i(p_2)(-i g_s)^2 \gamma_{\mu}  \frac{\Pi_{p_1}[n]}{(p_2+p_{12})^2} \gamma_{\mu} \frac{\slashed{p}_{1}+\slashed{p}_{2}+m_Q}{(p_1+p_2)^2-m_{Q}^{2}} \frac{-ie m_Q}{2 m_W \sin\theta_W} v_j(-p_3), \\
\mathcal{M}_{b}[n]&=&\mathcal{C} \bar{u}_i(p_2)(-i g_s)^2 \gamma_{\mu}  \frac{\Pi_{p_1}[n]}{(p_2+p_{12})^2} \frac{-ie m_Q}{2 m_W \sin\theta_W}  \frac{-\slashed{p}_{12}-\slashed{p}_{2}-\slashed{p}_{3}+m_Q}{(p_{12}+p_2+p_3)^2-m_{Q}^{2}} \gamma_{\mu} v_j(-p_3),  \\
\mathcal{M}_{c}[n]&=&\mathcal{C} \bar{u}_i(p_2)(-i g_s)^2 \gamma_{\mu}  \frac{\slashed{p}_{11}+\slashed{p}_{2}+\slashed{p}_{3}+m_{Q^{\prime}}}{(p_{11}+p_2+p_3)^2-m_{Q^{\prime}}^{2}} \frac{-ie m_{Q^{\prime}}}{2 m_W \sin\theta_W} \frac{\Pi_{p_1}[n]}{(p_3+p_{11})^2} \gamma_{\mu} v_j(-p_3), \\
\mathcal{M}_{d}[n]&=&\mathcal{C} \bar{u}_i(p_2)(-i g_s)^2 \frac{-ie m_{Q^{\prime}}}{2 m_W \sin\theta_W}  \frac{-\slashed{p}_{1}-\slashed{p}_{3}+m_{Q^{\prime}}}{(p_{1}+p_3)^2-m_{Q^{\prime}}^{2}}
\gamma_{\mu}  \frac{\Pi_{p_1}[n]}{(p_3+p_{11})^2} \gamma_{\mu} v_j(-p_3),\\
\mathcal{M}_{e}[n]&=& \frac{ie m_{W}\mathcal{C}}{\cos\theta_W^2 \sin\theta_W}   \frac{\bar{u}_i(p_2)\Gamma_{ZQ^{\prime}\bar{Q}^{\prime}}\Pi_{p_1}[n] \Gamma_{ZQ\bar{Q}}v_j(-p_3)}{((p_{12}+p_2)^2-m_{Z}^{2}+ i m_{Z} \Gamma_Z)
((p_{11}+p_3)^2-m_{Z}^{2}+ i m_{Z} \Gamma_Z)},  \label{gammazqq}\\
\mathcal{M}_{f}[n]&=&\int \frac{d ^4 l}{(2\pi)^4}\bar{u}_i(p_2)(-i g_s)^4 \gamma_{\mu}\frac{\Pi_{p_1}[n]}{(p_{12}+p_2)^2(p_{11}+p_3)^2}
\gamma_{\nu}v_j(-p_3) \label{loop1}\\
&&Tr\left[\frac{\slashed{l}-\slashed{p}_{1}-\slashed{p}_{2}-\slashed{p}_{3}+m_{q}}{(l-p_{1}-p_{2}-p_3)^2-m_{q}^{2}}
\gamma_{\mu}\frac{\slashed{l}-\slashed{p}_{11}-\slashed{p}_{3}+m_{q}}{(l-p_{11}-p_3)^2-m_{q}^{2}}
\gamma_{\nu}\frac{\slashed{l}+m_{q}}{l^2-m_{q}^{2}}\frac{-ie m_q\mathcal{C}}{2 m_W \sin\theta_W}\right],    \nonumber\\
\mathcal{M}_{g}[n]&=&\int \frac{d ^4 l}{(2\pi)^4}\bar{u}_i(p_2)(-i g_s)^4 \gamma_{\mu}\frac{\Pi_{p_1}[n]}{(p_{12}+p_2)^2(p_{11}+p_3)^2}
\gamma_{\nu}v_j(-p_3) \label{loop2}\\
&&Tr\left[\frac{\slashed{l}-\slashed{p}_{1}-\slashed{p}_{2}-\slashed{p}_3+m_{q}}{(l-p_{1}-p_{2}-p_3)^2-m_{q}^{2}}
\gamma_{\nu}\frac{\slashed{l}-\slashed{p}_{12}-\slashed{p}_{2}+m_{q}}{(l-p_{12}-p_2)^2-m_{q}^{2}}
\gamma_{\mu}\frac{\slashed{l}+m_{q}}{l^2-m_{q}^{2}}\frac{-ie m_q\mathcal{C}}{2 m_W \sin\theta_W}\right],    \nonumber
\end{eqnarray}
\end{widetext}
in which $\mathcal{C}$ stands for the color factor $\mathcal{C}_{ij,k}$, which will be described in detail in Sect~\ref{colorf}; $\theta_W$ is the Weinberg angle; the projector $\Pi_{p_1}[n]$ has the form of~\cite{Bodwin:2002hg}
\begin{widetext}
\begin{eqnarray}
\Pi_{p_1}[n] &=& \frac{1}{2\sqrt{M_{QQ^{\prime}}}}\epsilon[n](\slashed{p}_{1}+ M_{QQ^{\prime}}),
\end{eqnarray}
\end{widetext}
where $\epsilon[^1S_0]=\gamma_5$ and $\epsilon[^3S_1]=\slashed{\epsilon}$ with $\epsilon^\alpha$ is the polarization vector of the $^3S_1$ diquark state. $M_{QQ^{\prime}}\simeq M_{Q}+M_{Q^{\prime}}$ is adopted to ensure gauge invariance; $p_{11}$ and $p_{12}$ are the specific momenta of these two constituent quarks of the diquark state:
\begin{eqnarray}
p_{11}=\frac{m_{Q}}{M_{QQ^{\prime}}}p_{1}+p ~~\rm{and}~~\it p_{\rm{12}}=\frac{m_{Q^{\prime}}}{M_{QQ^{\prime}}}p_{\rm 1}-p,
\end{eqnarray}
where $p$ is the relative momentum between these two constituent quarks and it is small enough to neglect in the amplitude of $S$-wave state for the non-relativistic approximation.
In Eq.~(\ref{gammazqq}), $\Gamma_{ZQ\bar{Q}}$ and $\Gamma_{ZQ^{\prime}\bar{Q}^{\prime}}$ stand for the vertex of $Z^0$ boson with quark-antiquark pairs. Because the couplings for the $Z^0$ boson with the $b\bar{b}$ and $c\bar{c}$ pair are different, we do not state it definitely for the production of $\Xi_{cc}$, $\Xi_{bc}$ and $\Xi_{bb}$. In Eqs.~(\ref{loop1}) and (\ref{loop2}), $l$ is the loop momentum that needs to be integrated.

\subsection{Color factor} \label{colorf}

Given the different topologies in Fig.~\ref{diagram1}, four considered channels have different color structures, and we would like to take the channel $H^0 \rightarrow Q\bar{Q}$ as an example to explain how the color factor $\mathcal{C}_{ij,k}$ is calculated,
\begin{eqnarray}
\mathcal{C}_{ij,k}=\mathcal{N} \times \sum_{a,m,n} (T^a)_{im} (T^a)_{jn} \times G_{mnk},
\end{eqnarray}
where $i, j, m, n= 1, 2, 3$ are the color indices of the outgoing antiquarks $\bar{Q^{\prime}}$, $\bar{Q}$ and the two constituent quarks $Q$ and $Q^{\prime}$ of the diquark, respectively; $a=1, \ldots, 8$ and $k$ denote the color indices of the gluon and the diquark state $\langle QQ^{\prime}\rangle[n]$; the normalization constant $\mathcal{N}=\sqrt{1/2}$.
For the color-antitriplet $\bar 3$ state, the function $G_{mnk}$ is equal to the antisymmetric function $\varepsilon_{mnk}$, while will be the symmetric function $f_{mnk}$ for the color-sextuplet 6 state. The function $\varepsilon_{mnk}$ and $f_{mnk}$ satisfies
\begin{eqnarray}
\varepsilon_{mnk} \varepsilon_{m^{\prime}n^{\prime}k}=\delta_{mm^{\prime}}\delta_{nn^{\prime}}-\delta_{mn^{\prime}}\delta_{nm^{\prime}},
\nonumber\\f_{mnk} f_{m^{\prime}n^{\prime}k}=\delta_{mm^{\prime}}\delta_{nn^{\prime}}+\delta_{mn^{\prime}}\delta_{nm^{\prime}}.
\end{eqnarray}

After squaring the amplitude through $H^0 \rightarrow Q\bar{Q}$, the final color factor $\mathcal{C}^{2}_{ij,k}$ equals $\frac{4}{3}$ for the production of the color-antitriplet diquark state and $\frac{2}{3}$ for the color-sextuplet diquark state. Due to the different color matrices in the subgraphs of Fig.~\ref{diagram1}, the explicit color factors $\mathcal{C}^{2}_{ij,k}$ accompanying by the other two different channels are listed in Table~\ref{color}. Cross~term~1 stands for the cross term between $H^0 \rightarrow Q\bar{Q} / Q^{\prime} \bar{Q^{\prime}}$ and $H^0 \rightarrow Z^{0}Z^{0}$; Cross~term~2 is the cross term between $H^0 \rightarrow Q\bar{Q} / Q^{\prime} \bar{Q^{\prime}}$ and $H^0 \rightarrow gg$.
\begin{table}[htb]
\begin{center}
\caption{The color factors $\mathcal{C}^{2}_{ij,k}$ for different channels of Fig.~\ref{diagram1}.}
\begin{tabular}{c|c|c|c|c|c}
  \hline
  $C^{2}_{ij,k}$ & ~$H^0 \rightarrow Q\bar{Q}/Q^{\prime}\bar{Q^{\prime}}$~ & ~$H^0 \rightarrow Z^0 Z^0$~ & ~$H^0 \rightarrow gg$~ & ~Cross~term~1~ & ~Cross~term~2~ \\
  \hline
  color-antitriplet $\bar{3}$ &4/3& 3 & 1/3 & -2 & 2/3 \\
  color-sextuplet 6    &2/3& 6 & 1/6 &2& 1/3\\
  \hline
\end{tabular}
\label{color}
\end{center}
\end{table}

\section{Numerical results}

In numerical calculation, the input parameters are taken as follows \cite{Baranov:1995rc,Patrignani:2016xqp}:
\begin{eqnarray}
&&m_c=1.8~\rm{GeV},~~~\it{m_b}=\rm 5.1~{GeV},~~~\it{m_t}=\rm 173.0~{GeV},\nonumber\\
&&m_Z=91.1876~\rm{GeV},~~~\it{m_W}=\rm 80.385~{GeV},~~~\it{m_H}=\rm 125.18~{GeV}, \nonumber\\
&&M_{\Xi_{cc}}=3.6~{\rm GeV},~~~M_{\Xi_{bc}}=6.9~{\rm GeV},~~~M_{\Xi_{bb}}=10.2~{\rm GeV},\nonumber\\
&&|\Psi_{cc}(0)|^2=0.039~{\rm GeV}^3,~|\Psi_{bc}(0)|^2=0.065~{\rm GeV}^3,~|\Psi_{bb}(0)|^2=0.152~{\rm GeV}^3,\nonumber\\
&&\Gamma_{Z}=2.4952~{\rm GeV},~~~G_{F}=1.1663787 \times 10^{-5},
\end{eqnarray}
where the quark masses and wave functions are consistent with Ref.~\cite{Baranov:1995rc} and the others can be obtained from the PDG~\cite{Patrignani:2016xqp}.

We use FeynArts 3.9~\cite{Hahn:2000kx} to generate the amplitudes and the modified FormCalc 7.3/Loop-Tools 2.1~\cite{Hahn:1998yk} to do the algebraic and numerical calculations. The renormalization scale $\mu_r$ is set to be $2m_c$, $2m_c$ and $2m_b$ for the production of $\Xi_{cc}$, $\Xi_{bc}$ and $\Xi_{bb}$ correspondingly. Due to the total decay width of the Higgs boson not having been detected so accurately by the experiment, we consider the total decay width of the Higgs boson as 4.2~MeV~\cite{Heinemeyer:2013tqa} to estimate the branching ratio and corresponding events for the production of baryons $\Xi_{cc}$, $\Xi_{bc}$ and $\Xi_{bb}$.

\subsection{Basic results}

\begin{table}[htb]
\begin{center}
\caption{The decay widths for the process $H^0 \rightarrow b\bar{b} / c\bar{c}/Z^{0}Z^{0} / gg  \rightarrow \Xi_{QQ^{\prime}}+ \bar {Q^{\prime}} + \bar{Q}$, where $Q$ and $Q^{\prime}$ denote the heavy $c$ or $b$ quark. Cross~term~1 stands for the cross term between $H^0 \rightarrow Q\bar{Q} / Q^{\prime} \bar{Q^{\prime}}$ and $H^0 \rightarrow Z^{0}Z^{0}$, and Cross~term~2 is the cross term between $H^0 \rightarrow Q\bar{Q} / Q^{\prime} \bar{Q^{\prime}}$ and $H^0 \rightarrow gg$.}\vspace{0.5cm}
\begin{tabular}{c|c|c|c|c|c|c|c|c}
\hline
\multirow{2}{*}{$\Gamma$~(GeV)}
 & \multicolumn{2}{c}{$\Xi_{cc}$} & \multicolumn{4}{|c}{$\Xi_{bc}$} &
\multicolumn{2}{|c}{$\Xi_{bb}$} \\
\cline{2-9}
& $[^3S_1]_{\bar{3}}$ & $[^1S_0]_{6}$ &  $[^3S_1]_{\bar{3}}$ & $[^3S_1]_{6}$ & $[^1S_0]_{\bar{3}}$ & $[^1S_0]_{6}$ & $[^3S_1]_{\bar{3}}$ & $[^1S_0]_{6}$ \\
\hline
$H^0\rightarrow b\bar{b}(\times 10^{-7})$ & $-$ & $-$ & 5.89 & 2.95 & 4.48 & 2.24 & 0.41 & 0.28 \\
$H^0\rightarrow c\bar{c}(\times 10^{-7})$ & 0.65 & 0.35 & $1.03\times 10^{-2}$ & $5.16\times 10^{-3}$ & $1.23\times 10^{-2}$ & $6.16\times 10^{-3}$ & $-$ & $-$ \\
$H^0\rightarrow Q\bar{Q} / Q^{\prime} \bar{Q^{\prime}}(\times 10^{-7})$ & 0.65 & 0.35 & 5.87 & 2.94 & 4.57 & 2.29 & 0.41 & 0.28 \\
$H^0\rightarrow Z^{0}Z^{0}(\times 10^{-10})$ & $0.82$ & $1.63$ & 4.25 & 8.50 & 4.32 & 8.64 & 0.16 & 1.09 \\
$H^0\rightarrow gg(\times 10^{-9})$ & 3.01 & 0.47 & $2.36$ & $1.18$ & $1.00$ & $0.50$ & $0.41$ & $0.11$ \\
Cross~term~1($\times 10^{-10}$) & 0.10 & -0.33 & 2.45 & -2.45 & -8.25 & 8.25 & -0.57 & -9.16 \\
Cross~term~2($\times 10^{-9}$)& 0.65 & 5.64 & 2.48 & 1.24 & 20.58 & 10.29 & 0.63 & 2.91 \\
\hline
\end{tabular}
\label{hbb}
\end{center}
\end{table}
Based on the parameters mentioned before, four main Higgs decay channels for the production of $\Xi_{QQ'}$ have been analyzed carefully, and the decay width of each channel is presented in Table~\ref{hbb}. From Table~\ref{hbb}, we find that:
\begin{itemize}
  \item The biggest decay channel for the production of $\Xi_{cc}$ ($\Xi_{bb}$) is $H^0 \rightarrow c\bar{c}$ ($H^0 \rightarrow b\bar{b}$). Meanwhile for the production of $\Xi_{bc}$, the decay width in each diquark state through $H^0 \rightarrow b\bar{b}$ is about two orders of magnitude larger than that through $H^0 \rightarrow c\bar{c}$ mainly for the Yukawa coupling.
  \item From the decay widths through $H^0 \rightarrow Q\bar{Q} / Q^{\prime} \bar{Q^{\prime}}$, it can be seen that the contribution of the cross term between $H^0 \rightarrow b\bar{b}$ and $H^0 \rightarrow c\bar{c}$ is positive for $[^1S_0]_{\bar{3}/6}$ states and negative for $[^3S_1]_{\bar{3}/6}$ states.
  \item The decay widths through $H^0 \rightarrow Z^0Z^0/gg$ channels are very small and only a few percent compared to that through $H^0 \rightarrow Q\bar{Q} / Q^{\prime} \bar{Q^{\prime}}$.
  \item The contributions of the cross term between $H^0 \rightarrow Q\bar{Q} / Q^{\prime} \bar{Q^{\prime}}$ and $H^0 \rightarrow VV~(V=Z^0, g)$ should also be taken into account and the decay width for the production of baryons $\Xi_{QQ'}$ from these two cross terms are also listed in Table~\ref{hbb}.
\end{itemize}

To estimate the events of doubly heavy baryons produced at the ``Higgs factories'', the total decay width of the Higgs boson is needed to obtain the branching ratio correspondingly. But so far, the total decay width of Higgs boson could not be measured so accurately by the experiment and there was only given an upper limit of $13$~MeV~\cite{Khachatryan:2016ctc}. Here the total decay width of the Higgs boson is considered to be $4.2$~MeV as suggested by Ref.~\cite{Heinemeyer:2013tqa}. Running at $\sqrt{s} = 14$~TeV with the integrated luminosity of $3~ab^{-1}$, HL-LHC could
produce $1.65 \times 10^8$ Higgs bosons per year~\cite{LHCHIGGS}; the CEPC, the same as the ILC, would generate more than one million Higgs particles, mainly depending on the energy and integrated luminosity in operation~\cite{CEPCStudyGroup:2018rmc,Simon:2012ik}. Under these conditions, we can estimate the produced events of $\Xi_{QQ'}$ at the HL-LHC and the CEPC/ILC, respectively. The decay width and the estimated events through these four Higgs decay channels are showed in Table~\ref{sum}. By summing up the contribution from each intermediate diquark state, the total decay width, the branching ratio and the corresponding estimated events of the doubly heavy baryons $\Xi_{QQ'}$ could be obtained, which are given in Table~\ref{total}.

\begin{table}[htb]
\begin{center}
\caption{The decay width and the estimated events through these four Higgs decay channels at the HL-LHC and the CEPC/ILC.} \vspace{0.5cm}
\begin{tabular}{c|c|c|c|c|c|c|c|c}
\hline
\multirow{2}{*}{Fock~states}
 & \multicolumn{2}{c}{$\Xi_{cc}$} & \multicolumn{4}{|c}{$\Xi_{bc}$} &
\multicolumn{2}{|c}{$\Xi_{bb}$} \\
\cline{2-9}
& $[^3S_1]_{\bar{3}}$ & $[^1S_0]_{6}$ &  $[^3S_1]_{\bar{3}}$ & $[^3S_1]_{6}$ & $[^1S_0]_{\bar{3}}$ & $[^1S_0]_{6}$ & $[^3S_1]_{\bar{3}}$ & $[^1S_0]_{6}$ \\
\hline
$\Gamma~(\times 10^{-7}$GeV) & $0.69$ & $0.41$ & 5.93 & 2.97 & 4.78 & 2.41 & 0.42 & 0.30 \\
HL-LHC events ($\times 10^{4}$)& $0.27$ & $0.16$ & 2.33 & 1.17 & 1.88 & 0.95 & 0.17 & 0.12 \\
CEPC/ILC events ($\times 10^{2}$)& $0.16$ & $0.10$ & 1.41 & 0.71 & 1.14 & 0.57 & 0.10 & 0.07 \\
\hline
\end{tabular}
\label{sum}
\end{center}
\end{table}

\begin{table}[htb]
\begin{center}
\caption{The total decay width, the branching ratio and the estimated events of the doubly heavy baryons $\Xi_{QQ'}$ by summing up the contribution from each intermediate diquark state.} \vspace{0.5cm}
\begin{tabular}{c|c|c|c|c}
\hline
  & ~$\Gamma~(\times 10^{-7}~\rm{GeV})$~ &~Br $(\times 10^{-4})$~ &  ~HL-LHC events~ & ~CEPC/ILC events~ \\
\hline
$H^0  \rightarrow \Xi_{cc}$ & $1.10$ & 0.26 & $0.43\times 10^{4}$ & $0.26\times 10^{2}$ \\
$H^0  \rightarrow \Xi_{bc}$ & $16.09$ & 3.83 & $6.32\times 10^{4}$ & $3.83\times 10^{2}$ \\
$H^0  \rightarrow \Xi_{bb}$ & $0.72$ & 0.17 & $0.28\times 10^{4}$ & $0.17\times 10^{2}$ \\
\hline
\end{tabular}
\label{total}
\end{center}
\end{table}

Tables~\ref{sum} and~\ref{total} show that
\begin{itemize}
  \item The estimated events of $\Xi_{bc}$ are about one order of magnitude larger than that of $\Xi_{cc}$ and $\Xi_{bb}$.
  \item The branching ratio via Higgs boson decays is about $10^{-4}$ for the production of $\Xi_{bc}$ baryon, and $10^{-5}$ for the production of $\Xi_{cc}$ and $\Xi_{bb}$ baryons.
  \item At the HL-LHC, there are sizable events of doubly heavy baryons $\Xi_{QQ'}$, at the order of $10^4$, produced per year.
  \item There are only about $10^2$ $\Xi_{QQ'}$ events produced at the CEPC/ILC, but with a cleaner background. In view of the upgrade of the CEPC/ILC, such as increasing the luminosity to the same level as the HL-LHC, there would be 3.75 times the $\Xi_{QQ'}$ events.
\end{itemize}

\begin{figure}[htb]
  \centering
  \includegraphics[width=0.33\textwidth]{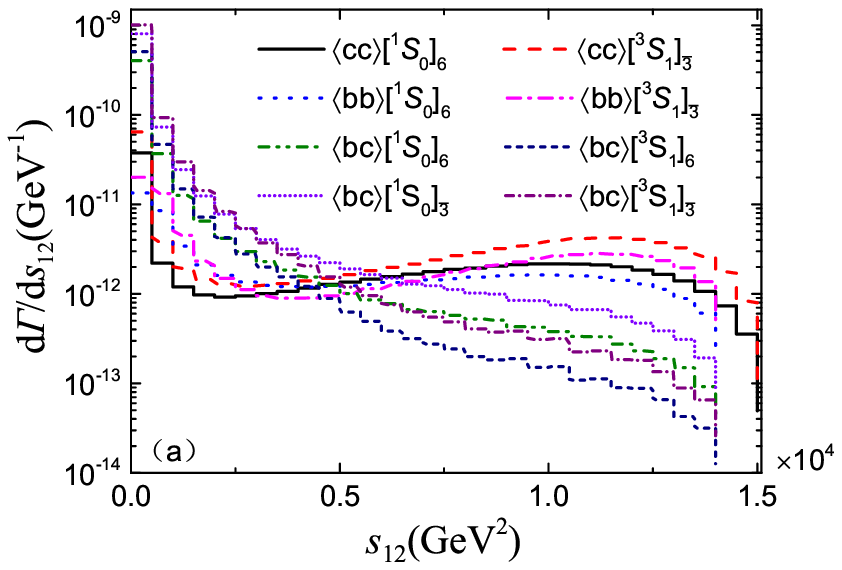}
  \includegraphics[width=0.33\textwidth]{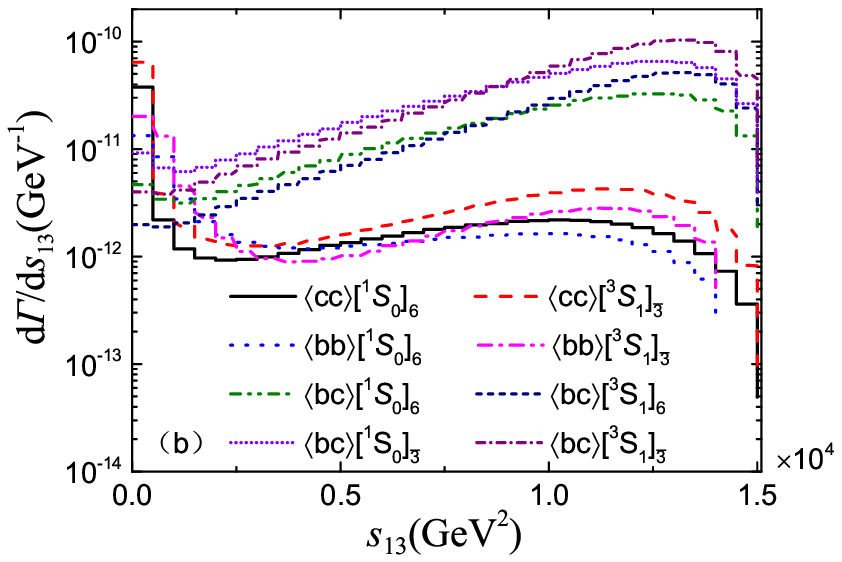}
  \includegraphics[width=0.32\textwidth]{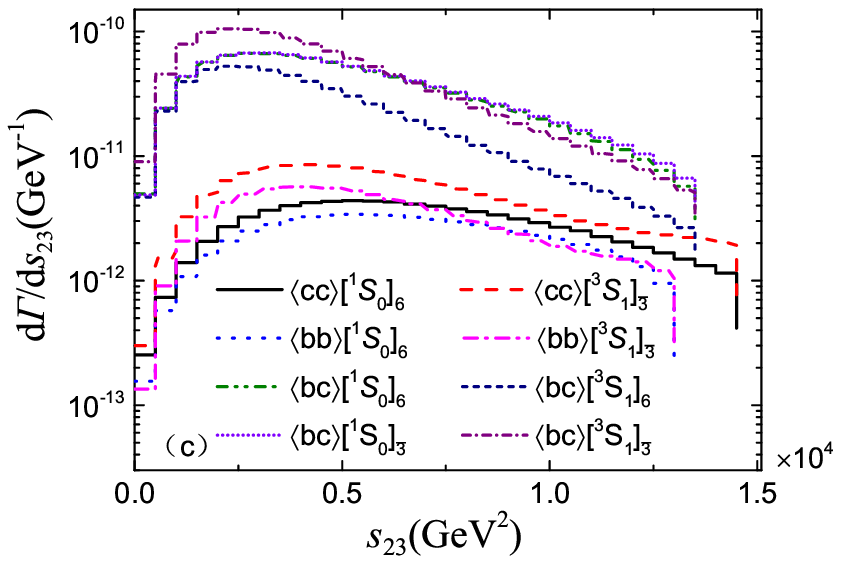}
  \caption{The invariant mass differential decay widths $d\Gamma/ds_{12}$ (a), $d\Gamma/ds_{13}$ (b) and $d\Gamma/ds_{23}$ (c) for the process $H^{0} \rightarrow \Xi_{QQ^{\prime}}+ \bar {Q^{\prime}} + \bar {Q}$. Eight lines represent the possible spin and color configurations, i.e.,
$\langle cc\rangle[^{1}S_{0}]_{\mathbf{6}}$, $\langle cc\rangle[^{3}S_{1}]_{\mathbf{\bar 3}}$, $\langle bc\rangle[^{3}S_{1}]_{\mathbf{\bar 3}/ \mathbf{6}}$, $\langle bc\rangle[^{1}S_{0}]_{\mathbf{\bar 3}/ \mathbf{6}}$, $\langle bb\rangle[^{1}S_{0}]_{\mathbf{6}}$ and $\langle bb\rangle[^{3}S_{1}]_{\mathbf{\bar 3}}$.}
  \label{totals} 
\end{figure}

\begin{figure}[htb]
  \centering
  \includegraphics[width=0.33\textwidth]{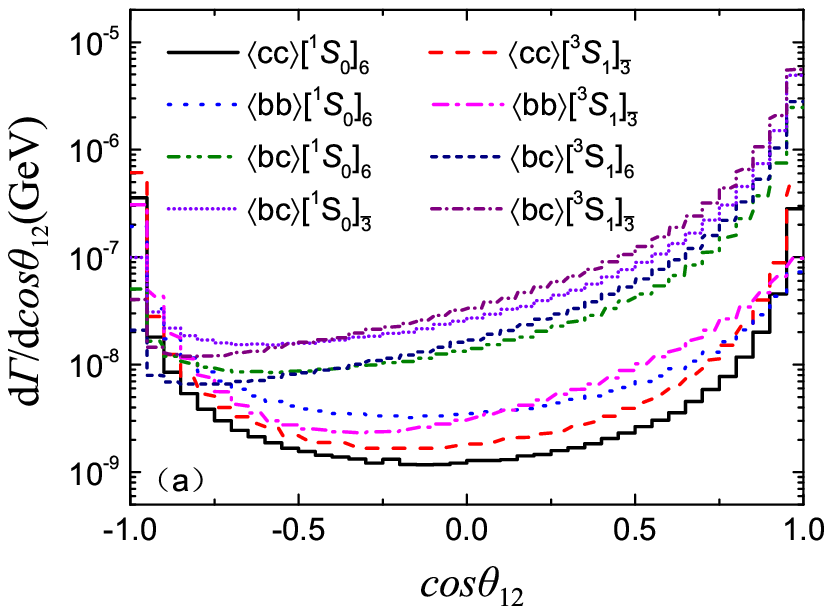}
  \includegraphics[width=0.33\textwidth]{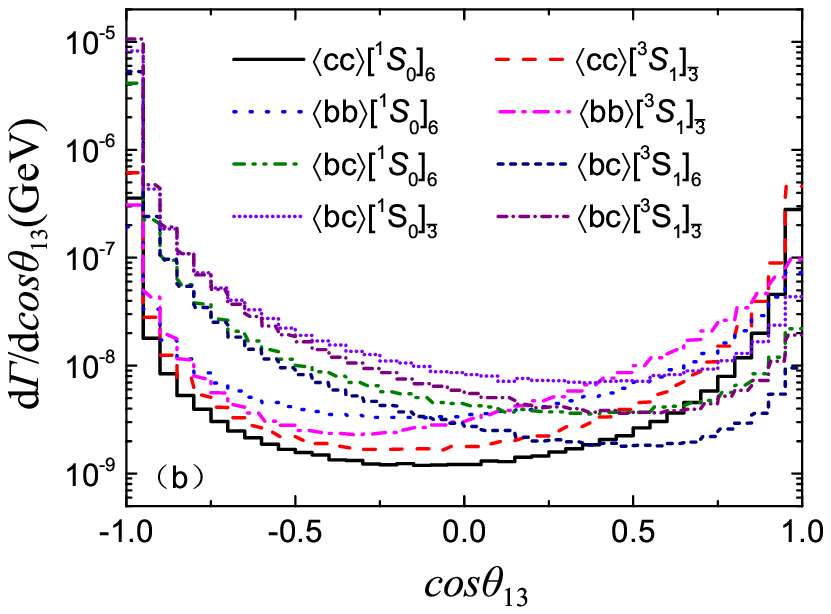}
  \includegraphics[width=0.32\textwidth]{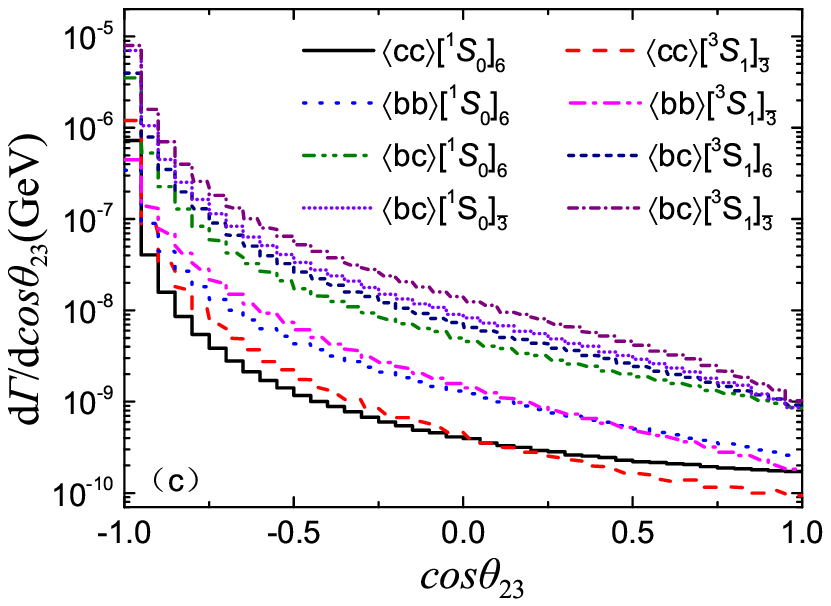}
  \caption{The angular differential decay widths $d\Gamma/dcos\theta_{12}$ (a), $d\Gamma/dcos\theta_{13}$ (b) and $d\Gamma/dcos\theta_{23}$ (c) for the process $H^{0} \rightarrow \Xi_{QQ^{\prime}} + \bar {Q^{\prime}} + \bar {Q}$. Eight lines represent the possible spin and color configurations, i.e.,
$\langle cc\rangle[^{1}S_{0}]_{\mathbf{6}}$, $\langle cc\rangle[^{3}S_{1}]_{\mathbf{\bar 3}}$, $\langle bc\rangle[^{3}S_{1}]_{\mathbf{\bar 3}/ \mathbf{6}}$, $\langle bc\rangle[^{1}S_{0}]_{\mathbf{\bar 3}/ \mathbf{6}}$, $\langle bb\rangle[^{1}S_{0}]_{\mathbf{6}}$ and $\langle bb\rangle[^{3}S_{1}]_{\mathbf{\bar 3}}$.}
  \label{totalcos} 
\end{figure}

To make a clear analysis of the distributions for the production of $\Xi_{QQ'}$ through these four considered channels and to be helpful as regards experimental detection, the invariant mass differential decay widths $d\Gamma/ds_{ij}$ and the angular differential distributions $d\Gamma/dcos\theta_{ij}$ are plotted in Figs.~\ref{totals} and \ref{totalcos}, where the invariant mass $s_{ij}=(p_{i}+p_{j})^2$ and $\theta_{ij}$ is the angle between the momenta $\overrightarrow{p_{i}}$ and $\overrightarrow{p_{j}}$ in the Higgs boson rest frame.
All the possible spin and color configurations have been taken into consideration, i.e.,
$\langle cc\rangle[^{1}S_{0}]_{\mathbf{6}}$, $\langle cc\rangle[^{3}S_{1}]_{\mathbf{\bar 3}}$, $\langle bc\rangle[^{3}S_{1}]_{\mathbf{\bar 3}/ \mathbf{6}}$, $\langle bc\rangle[^{1}S_{0}]_{\mathbf{\bar 3}/ \mathbf{6}}$, $\langle bb\rangle[^{1}S_{0}]_{\mathbf{6}}$ and $\langle bb\rangle[^{3}S_{1}]_{\mathbf{\bar 3}}$.

Figures~\ref{totals} and \ref{totalcos} show that the behaviors of the differential distributions for the production of baryon $\Xi_{bc}$ are different from that for the production of baryons $\Xi_{cc}$ and $\Xi_{bb}$. Figure~\ref{totals}a shows that, as $s_{12}$ gets smaller and smaller, i.e., $p_1$ and  $p_2$ are collinear, there is a maximal value of $d\Gamma/ds_{12}$. From Fig.~\ref{totalcos}a (3b), one finds that, for the production of $\Xi_{bc}$, $d\Gamma/dcos\theta_{12}$ ($d\Gamma/dcos\theta_{13}$) is seen to be the largest when $\cos\theta_{12}=1$ ($\cos\theta_{13}=-1$), i.e., the doubly heavy baryons $\Xi_{QQ'}$ and the heavy quark $\bar{Q'}$ ($\bar{Q}$) move side by side (back to back). Figure~\ref{totalcos}c illustrates this fact again. Meanwhile for the production of $\Xi_{cc}$ and $\Xi_{bb}$, there are similar kinematic behaviors for (a) and (b) in Figs.~\ref{totals} and~\ref{totalcos} for the identical particles in the diquark state.

Concerning the discovery potential of these baryons at the HL-LHC and CEPC/ILC, the possible decay channels of $\Xi_{QQ'}$ is useful. Similar to the observation of $\Xi_{cc}^{++}$ baryon, the $\Xi_{bc}$ and $\Xi_{bb}$ baryons could be observed by cascade decays such as $\Xi_{bc}^{+}\to\Xi_{cc}^{++}(\to p K^-\pi^+ \pi^+)\pi^{-}$ and $\Xi^{0}_{bb}\to\Xi_{bc}^+ (\to\Xi^{++}\pi^-) \pi^-$.
At present, many phenomenological models have been suggested to study the decay properties of the doubly heavy baryons, which are at the initial stage for the large non-perturbative effects. An overview of the doubly heavy baryons decay, together with the possibilities of observation may be found in Refs.\cite{Bediaga:2012py, Bediaga:2018lhg}. As for the detection efficiency in the experiment, the events cannot be 100$\%$ detected. Compared to the $\Xi_{cc}^{++}$ events detected by LHCb \cite{Aaij:2017ueg,Chang:2006eu,Chang:2007pp,Chang:2009va}, about $\mathcal{O}(10)$ doubly heavy baryons $\Xi_{QQ'}$ events from Higgs boson decays would be detected per year at the HL-LHC and there would be $\mathcal{O}(1)$ events which could be detected at the ILC and CEPC.

\subsection{Theoretical uncertainties}

In this subsection, the theoretical uncertainties for the production of $\Xi_{QQ'}$ via the Higgs boson decays would be discussed. There are three main sources of the theoretical uncertainties: the quark mass, the renormalization scale $\mu_r$ and the transition probability. The likely quark mass uncertainty covers $m_c$ and $m_b$ for building the mass of the corresponding doubly heavy baryons $\Xi_{QQ^{\prime}}$. We shall analyze the caused quark mass uncertainties by varying $m_c=1.8\pm0.3~\rm{GeV}$ and $m_b=5.1\pm0.4~\rm{GeV}$, which are listed in Table~\ref{mc} and \ref{mb}, respectively. It is worth mentioning that, discussing the uncertainty caused by one parameter, the others should be fixed to their central values. The decay widths through these four considered decay channels have been summed up for the total decay width. Table~\ref{mc} and \ref{mb} show that the decay width for the production of $\Xi_{cc}$ ($\Xi_{bb}$) decreases with the increment of $m_c$ ($m_b$) for a suppression of the phase space. Meanwhile, for the production of $\Xi_{bc}$, the decay width decreases with the increment of $m_c$ but increases with the increment of $m_b$.

\begin{table}[htb]
\begin{center}
\caption{The theoretical uncertainties for the production of baryons $\Xi_{QQ^{\prime}}$ via Higgs boson decays by varying $m_c=1.8 \pm 0.3~\rm{GeV}$.} \vspace{0.5cm}
\begin{tabular}{c|c|c|c|c|c|c|c|c}
\hline
\multirow{2}{*}{$\Gamma (\times 10^{-7}$GeV)}
 & \multicolumn{2}{c}{$\Xi_{cc}$} & \multicolumn{4}{|c}{$\Xi_{bc}$} &
\multicolumn{2}{|c}{$\Xi_{bb}$} \\
\cline{2-9}
& $[^3S_1]_{\bar{3}}$ & $[^1S_0]_{6}$ &  $[^3S_1]_{\bar{3}}$ & $[^3S_1]_{6}$ & $[^1S_0]_{\bar{3}}$ & $[^1S_0]_{6}$ & $[^3S_1]_{\bar{3}}$ & $[^1S_0]_{6}$ \\
\hline
$m_c$=1.5~GeV & $0.85$ & $0.49$ & 10.90 & 5.46 & 8.17 & 4.11 & 0.42 & 0.30 \\
$m_c$=1.8~GeV & $0.69$ & $0.41$ & 5.93 & 2.97 & 4.78 & 2.41 & 0.42 & 0.30 \\
$m_c$=2.1~GeV & $0.58$ & $0.35$ & 3.54 & 1.77 & 3.07 & 1.55 & 0.42 & 0.30 \\
\hline
\end{tabular}
\label{mc}
\end{center}
\end{table}

\begin{table}[htb]
\begin{center}
\caption{The theoretical uncertainties for the production of baryons $\Xi_{QQ^{\prime}}$ via Higgs boson decays by varying $m_b=5.1 \pm 0.4~\rm{GeV}$.} \vspace{0.5cm}
\begin{tabular}{c|c|c|c|c|c|c|c|c}
\hline
\multirow{2}{*}{$\Gamma (\times 10^{-7}$GeV)}
 & \multicolumn{2}{c}{$\Xi_{cc}$} & \multicolumn{4}{|c}{$\Xi_{bc}$} &
\multicolumn{2}{|c}{$\Xi_{bb}$} \\
\cline{2-9}
& $[^3S_1]_{\bar{3}}$ & $[^1S_0]_{6}$ &  $[^3S_1]_{\bar{3}}$ & $[^3S_1]_{6}$ & $[^1S_0]_{\bar{3}}$ & $[^1S_0]_{6}$ & $[^3S_1]_{\bar{3}}$ & $[^1S_0]_{6}$ \\
\hline
$m_b$=4.7~GeV & $0.69$ & $0.41$ & 4.94 & 2.47 & 4.11 & 2.07 & 0.46 & 0.33 \\
$m_b$=5.1~GeV& $0.69$ & $0.41$ & 5.93 & 2.97 & 4.78 & 2.41 & 0.42 & 0.30 \\
$m_b$=5.5~GeV& $0.69$ & $0.41$ & 7.02 & 3.51 & 5.51 & 2.78 & 0.38 & 0.28 \\
\hline
\end{tabular}
\label{mb}
\end{center}
\end{table}

Due to the QCD running coupling, the renormalization scale $\mu_r$ would make a significant contribution to the decay width. We could obtain the uncertainties by substituting three different renormalization scales, i.e., $\mu_r=2m_c$, $\rm{M}_{bc}$ or $2m_b$, which are presented in Table~\ref{mqun}. As a supplement, the QCD running coupling $\alpha_s(\mu_r)$ is also added to Table~\ref{mqun}. In fact, such scale ambiguity could be suppressed by a higher-order perturbative calculation or proper scale-setting methods such as the Principle of Maximum Conformal (PMC) method \cite{Brodsky:2013vpa,Brodsky:2012rj,Mojaza:2012mf,Brodsky:2011ta}.
\begin{table}[htb]
\begin{center}
\caption{The theoretical uncertainties for the production of baryons $\Xi_{QQ^{\prime}}$ via Higgs boson decays by substituting the renormalization scale $\mu_r=2m_c$, $\rm{M}_{bc}$ or $2m_b$. The units of the decay widths are $(\times 10^{-7}$~GeV).}
\begin{tabular}{cccc}
  \hline
  $\mu_r$ & ~$2m_c$~ & $~\rm{M}_{bc}~$ & ~2$m_b$~ \\
  $\alpha_s$ & ~0.242~ &~0.198~ & ~0.180~ \\
  \hline
  $\Gamma_{\Xi_{cc}[^3 S_1]_{\bar{3}}}$ &$0.69$& 0.45 & 0.37 \\
  $\Gamma_{\Xi_{cc}[^1 S_0]_{6}}$       &$0.41$ & 0.27 & 0.22 \\
  $\Gamma_{\Xi_{bc}[^3 S_1]_{\bar{3}}}$ &5.93   &3.94  &3.24   \\
  $\Gamma_{\Xi_{bc}[^3 S_1]_{6}}      $ &2.97  &1.97  & 1.63  \\
  $\Gamma_{\Xi_{bc}[^1 S_0]_{\bar{3}}}$ &4.78  &3.16  & 2.59  \\
  $\Gamma_{\Xi_{bc}[^1 S_0]_{6}}$       &2.41  &1.60 & 1.31  \\
  $\Gamma_{\Xi_{bb}[^3 S_1]_{\bar{3}}}$ &0.77 & 0.51 & 0.42  \\
  $\Gamma_{\Xi_{bb}[^1 S_0]_{6}}$       &0.57 & 0.37 & 0.30 \\
  \hline
\end{tabular}
\label{mqun}
\end{center}
\end{table}

Finally, the theoretical uncertainty caused by the non-perturbative transition probability is considered. Considering that the transition probability is proportional to the decay width, its uncertainty can be conventionally obtained when we know its exact value. Throughout the paper, the transition probability of the color-antitriplet diquark state $\langle QQ^{\prime}\rangle_{\bar{3}}$ and the color-sextuplet diquark state $\langle QQ^{\prime}\rangle_6$ to the heavy baryon $\Xi_{QQ^{\prime}}$ have been considered as the same, i.e., $h_{6} \simeq h_{\bar 3}=|\Psi_{QQ^{\prime}}(0)|^2$ \cite{Bagan:1994dy,Petrelli:1997ge}, where the wave functions at the origin $|\Psi_{QQ^{\prime}}(0)|$ are derived from the power-law potential model. However, there is a larger uncertainty for $h_{6}$ than for $h_{\bar 3}$. Within the framework of NRQCD, the intermediate diquark state $\langle QQ^{\prime}\rangle[n]$ can be expanded into a series of Fock states with the relative velocity ($v$) and, according to the NRQCD power counting rule, each Fock state is of the same importance, which is the main reason why we took $h_{6} \simeq h_{\bar 3}$. In addition to this point of view, there is another point of view: namely, that the color-sextuplet state would be suppressed by $v^2$ compared to the color-antitriplet state, i.e., $h_{6}/v^2 \simeq h_{\bar 3}=|\Psi_{QQ^{\prime}}(0)|^2$. Even if the contribution of the color-sextuplet diquark $(QQ^{\prime})_6$ state can be ignored ($h_{6}=0$) and only the color-antitriplet diquark $(QQ^{\prime})_{\bar{3}}$ state is taken into consideration ($h_{\bar 3}=|\Psi_{QQ^{\prime}}(0)|^2$), there are still 0.27$\times 10^{4}$ events of $\Xi_{cc}$, 4.21 $\times 10^{4}$ events of $\Xi_{bc}$ and 0.17 $\times 10^{4}$ events of $\Xi_{bb}$ produced per year at the HL-LHC. However, there are fewer events produced at the CEPC/ILC, only 0.16$\times 10^{2}$ events of $\Xi_{cc}$, 2.55 $\times 10^{2}$ events of $\Xi_{bc}$ and 0.10 $\times 10^{2}$ events of $\Xi_{bb}$.

\section{Summary}

Within the framework of NRQCD, the decay widths for the production of baryons $\Xi_{cc}$, $\Xi_{bc}$ and $\Xi_{bb}$ have been analyzed through four main Higgs decay channels, $H^0 \rightarrow b\bar{b} / c\bar{c}/Z^{0}Z^{0} / gg$. By summing up all
the contributions from the intermediate diquark states, $\langle cc\rangle[^{1}S_{0}]_{\mathbf{6}}$, $\langle cc\rangle[^{3}S_{1}]_{\mathbf{\bar 3}}$, $\langle bc\rangle[^{3}S_{1}]_{\mathbf{\bar 3}/ \mathbf{6}}$, $\langle bc\rangle[^{1}S_{0}]_{\mathbf{\bar 3}/ \mathbf{6}}$, $\langle bb\rangle[^{1}S_{0}]_{\mathbf{6}}$ and $\langle bb\rangle[^{3}S_{1}]_{\mathbf{\bar 3}}$, the total decay width for the process $H^0 \rightarrow \Xi_{QQ^{\prime}} + \bar {Q^{\prime}} +\bar {Q}$ can be obtained, i.e.,
\begin{eqnarray}
&&\Gamma_{H^0 \rightarrow \Xi_{cc}}=1.10^{+0.24}_{-0.17}\times 10^{-7}~\rm{GeV},\nonumber \\
&&\Gamma_{H^0 \rightarrow \Xi_{bc}}=16.09^{+12.55}_{-6.16} \times 10^{-7}~\rm{GeV}, \nonumber\\
&&\Gamma_{H^0 \rightarrow \Xi_{bb}}=0.72^{+0.07}_{-0.06} \times 10^{-7}~\rm{GeV},
\nonumber
\end{eqnarray}
where the uncertainty is caused by varying the quark mass $m_c=1.8\pm0.3$~GeV and $m_b=5.1\pm0.4$~GeV. To be helpful as regards experimental detection, the invariant mass and the angular differential distributions have also been presented. The corresponding produced events of the doubly heavy baryons $\Xi_{QQ'}$ are both estimated at the HL-LHC and the CEPC/ILC. There are about $(0.27\sim0.43)\times 10^{4}$ events of $\Xi_{cc}$, $(4.21\sim6.32)\times 10^{4}$ events of $\Xi_{bc}$ and $(0.17\sim0.28)\times 10^{4}$ events of $\Xi_{bb}$ produced per year at the HL-LHC. There are fewer events produced at the CEPC/ILC, only about $(0.16\sim0.26)\times 10^{2}$ events of $\Xi_{cc}$, $(2.55\sim3.83)\times 10^{2}$ events of $\Xi_{bc}$ and $(0.10\sim0.17)\times 10^{2}$ events of $\Xi_{bb}$,
where the uncertainties are from the transition probability. Due to the high luminosity and high collision energy, there are sizable events of doubly heavy baryons $\Xi_{QQ^{\prime}}$ produced per year at the HL-LHC via Higgs boson decays, which will be accessible by experiment research.

\hspace{2cm}

{\bf Acknowledgements}: This work was partially supported by the National Natural Science Foundation of China (nos.11375008, 11647307, 11625520, 11847301). This research was also supported by Conselho Nacional de Desenvolvimento Cient\'{\i}fico e Tecnol\'ogico (CNPq),
and Coordena\c{c}\~ao de Aperfei\c{c}oamento de Pessoal de N\'ivel Superior (CAPES).


\begin{thebibliography}{99}

\bibitem{LHCHIGGS}
 LHC Higgs Cross Section Working Group, Higgs production cross se AN2(2016). https://twiki.cern.ch/twiki/bin/view/LHCPhysics
/HiggsEuropeanStrategy$\#$SM  

\bibitem{CEPCStudyGroup:2018rmc}
  [CEPC Study Group],
  CEPC Conceptual Design Report(2018).
  arXiv:1809.00285 [physics.acc-ph]

\bibitem{Simon:2012ik}
  F.~Simon,
  Prospects for Precision Higgs Physics at Linear Colliders,
  PoS ICHEP {\bf 2012}, 066 (2013)

\bibitem{Aad:2015sda}
  G.~Aad {\it et al.} [ATLAS Collaboration],
  Search for Higgs and Z Boson Decays to $J/\psi \gamma$ and $\Upsilon(nS)\gamma$ with the ATLAS Detector.
  Phys.\ Rev.\ Lett.\  {\bf 114}, 121801 (2015)

\bibitem{Achasov:1991ms}
  N.~N.~Achasov and V.~K.~Besprozvannykh,
  Decays $\psi, \Upsilon \rightarrow H (a) \gamma$ and $H \rightarrow \psi \gamma, \Upsilon \gamma$.
  Sov.\ J.\ Nucl.\ Phys.\  {\bf 55}, 1072 (1992)

\bibitem{Bodwin:2014bpa}
  G.~T.~Bodwin, H.~S.~Chung, J.~H.~Ee, J.~Lee and F.~Petriello,
  Relativistic corrections to Higgs boson decays to quarkonia.
  Phys.\ Rev.\ D {\bf 90}, 113010 (2014)

  G.~T.~Bodwin, F.~Petriello, S.~Stoynev and M.~Velasco,
  Higgs boson decays to quarkonia and the $H\bar{c}c$  coupling.
  Phys.\ Rev.\ D {\bf 88}, 053003 (2013)

\bibitem{Koenig:2015pha}
  M.~K$\ddot{\rm{o}}$nig and M.~Neubert,
  Exclusive Radiative Higgs Decays as Probes of Light-Quark Yukawa Couplings.
  JHEP {\bf 1508}, 012 (2015)

\bibitem{Qiao:1998kv}
  C.~F.~Qiao, F.~Yuan and K.~T.~Chao,
  Quarkonium production in SM Higgs decays.
  J.\ Phys.\ G {\bf 24}, 1219 (1998)

\bibitem{Jiang:2015pah}
  J.~Jiang and C.~F.~Qiao,
  $B_c$ Production in Higgs Boson Decays.
  Phys.\ Rev.\ D {\bf 93}, 054031 (2016)


\bibitem{Liao:2018nab}
  Q.~L.~Liao, Y.~Deng, Y.~Yu, G.~C.~Wang and G.~Y.~Xie,
  Heavy $P$-wave quarkonium production via Higgs decays.
  Phys.\ Rev.\ D {\bf 98}, 036014 (2018)

\bibitem{Aaij:2017ueg}
  R.~Aaij {\it et al.} [LHCb Collaboration],
  Observation of the doubly charmed baryon $\Xi_{cc}^{++}$.
  Phys.\ Rev.\ Lett.\  {\bf 119}, 112001 (2017)

\bibitem{GellMann:1964nj}
  M.~Gell-Mann,
  A Schematic Model of Baryons and Mesons.
  Phys.\ Lett.\  {\bf 8}, 214 (1964)

\bibitem{Zweig:1981pd}
  G.~Zweig,
  An SU(3) model for strong interaction symmetry and its breaking. Version 1,
  CERN-TH-401 (1964)

\bibitem{Zweig:1964jf}
  G.~Zweig,
  An SU(3) model for strong interaction symmetry and its breaking. Version 2, Developments in the Quark Theory of Hadrons, Volume 1. Edited by D. Lichtenberg and S. Rosen. pp. 22-101 (1980)

\bibitem{DeRujula:1975qlm}
  A.~De Rujula, H.~Georgi and S.~L.~Glashow,
  Hadron Masses in a Gauge Theory.
  Phys.\ Rev.\ D {\bf 12}, 147 (1975)

\bibitem{Bodwin:1994jh}
  G.~T.~Bodwin, E.~Braaten and G.~P.~Lepage,
  Rigorous QCD analysis of inclusive annihilation and production of heavy quarkonium.
  Phys.\ Rev.\ D {\bf 51}, 1125 (1995)

\bibitem{Petrelli:1997ge}
  A.~Petrelli, M.~Cacciari, M.~Greco, F.~Maltoni and M.~L.~Mangano,
  NLO production and decay of quarkonium.
  Nucl.\ Phys.\ B {\bf 514}, 245 (1998)

\bibitem{Kiselev:1994pu}
  V.~V.~Kiselev, A.~K.~Likhoded and M.~V.~Shevlyagin,
  Double charmed baryon production at B factory.
  Phys.\ Lett.\ B {\bf 332}, 411 (1994)

\bibitem{Ma:2003zk}
  J.~P.~Ma and Z.~G.~Si,
  Factorization approach for inclusive production of doubly heavy baryon.
  Phys.\ Lett.\ B {\bf 568}, 135 (2003)

\bibitem{Zheng:2015ixa}
  X.~C.~Zheng, C.~H.~Chang and Z.~Pan,
  Production of doubly heavy-flavored hadrons at $e^+e^-$ colliders.
  Phys.\ Rev.\ D {\bf 93}, 034019 (2016)

\bibitem{Jiang:2012jt}
  J.~Jiang, X.~G.~Wu, Q.~L.~Liao, X.~C.~Zheng and Z.~Y.~Fang,
  Doubly Heavy Baryon Production at A High Luminosity $e^+ e^-$ Collider.
  Phys.\ Rev.\ D {\bf 86}, 054021 (2012)

\bibitem{Berezhnoy:1995fy}
  A.~V.~Berezhnoy, V.~V.~Kiselev and A.~K.~Likhoded,
  Hadronic production of baryons containing two heavy quarks.
  Phys.\ Atom.\ Nucl.\  {\bf 59}, 870 (1996)

\bibitem{Doncheski:1995ye}
  M.~A.~Doncheski, J.~Steegborn and M.~L.~Stong,
  Fragmentation production of doubly heavy baryons.
  Phys.\ Rev.\ D {\bf 53}, 1247 (1996)

\bibitem{Baranov:1995rc}
  S.~P.~Baranov,
  On the production of doubly flavored baryons in p p, e p and gamma gamma collisions.
  Phys.\ Rev.\ D {\bf 54}, 3228 (1996)

\bibitem{Berezhnoy:1998aa}
  A.~V.~Berezhnoy, V.~V.~Kiselev, A.~K.~Likhoded and A.~I.~Onishchenko,
  Doubly charmed baryon production in hadronic experiments.
  Phys.\ Rev.\ D {\bf 57}, 4385 (1998)

\bibitem{Chang:2006eu}
  C.~H.~Chang, C.~F.~Qiao, J.~X.~Wang and X.~G.~Wu,
  Estimate of the hadronic production of the doubly charmed baryon $\Xi_{cc}$ under GM-VFN scheme.
  Phys.\ Rev.\ D {\bf 73}, 094022 (2006)

\bibitem{Chang:2007pp}
  C.~H.~Chang, J.~X.~Wang and X.~G.~Wu,
  GENXICC: A Generator for hadronic production of the double heavy baryons $\Xi_{cc}$, $\Xi_{bc}$ and $\Xi_{bb}$.
  Comput.\ Phys.\ Commun.\  {\bf 177}, 467 (2007)

\bibitem{Chang:2009va}
  C.~H.~Chang, J.~X.~Wang and X.~G.~Wu,
  GENXICC2.0: An Upgraded Version of the Generator for Hadronic Production of Double Heavy Baryons $\Xi_{cc}$, $\Xi_{bc}$ and $\Xi_{bb}$.
  Comput.\ Phys.\ Commun.\  {\bf 181}, 1144 (2010)

\bibitem{Zhang:2011hi}
  J.~W.~Zhang, X.~G.~Wu, T.~Zhong, Y.~Yu and Z.~Y.~Fang,
  Hadronic Production of the Doubly Heavy Baryon $\Xi_{bc}$ at LHC.
  Phys.\ Rev.\ D {\bf 83}, 034026 (2011)

\bibitem{Wang:2012vj}
  X.~Y.~Wang and X.~G.~Wu,
  GENXICC2.1: An Improved Version of GENXICC for Hadronic Production of Doubly Heavy Baryons.
  Comput.\ Phys.\ Commun.\  {\bf 184}, 1070 (2013)

\bibitem{Chen:2014hqa}
  G.~Chen, X.~G.~Wu, J.~W.~Zhang, H.~Y.~Han and H.~B.~Fu,
  Hadronic production of $\Xi_{cc}$ at a fixed-target experiment at the LHC.
  Phys.\ Rev.\ D {\bf 89}, 074020 (2014)

\bibitem{Li:2007vy}
  S.~Y.~Li, Z.~G.~Si and Z.~J.~Yang,
  Doubly heavy baryon production at gamma gamma collider.
  Phys.\ Lett.\ B {\bf 648}, 284 (2007)

\bibitem{Chen:2014frw}
  G.~Chen, X.~G.~Wu, Z.~Sun, Y.~Ma and H.~B.~Fu,
  Photoproduction of doubly heavy baryon at the ILC.
  JHEP {\bf 1412}, 018 (2014)

\bibitem{Huan-Yu:2017emk}
  H.~Y.~Bi, R.~Y.~Zhang, X.~G.~Wu, W.~G.~Ma, X.~Z.~Li and S.~Owusu,
  Photoproduction of doubly heavy baryon at the LHeC.
  Phys.\ Rev.\ D {\bf 95}, 074020 (2017)

\bibitem{Yao:2018zze}
  X.~Yao and B.~M$\ddot{u}$ller,
  Doubly charmed baryon production in heavy ion collisions.
  Phys.\ Rev.\ D {\bf 97}, 074003 (2018)

\bibitem{Chen:2018koh}
  G.~Chen, C.~H.~Chang and X.~G.~Wu,
  Hadronic production of the doubly charmed baryon via the proton–nucleus and the nucleus–nucleus collisions at the RHIC and LHC.
  Eur.\ Phys.\ J.\ C {\bf 78}, 801 (2018)

\bibitem{topdecay}
  J.~J.~Niu, L.~Guo, H.~H.~Ma and X.~G.~Wu,
  Production of semi-inclusive doubly heavy baryon via top quark decays.
  Phys.\ Rev.\ D {\bf 98}, 094021 (2018)

\bibitem{Patrignani:2016xqp}
  M.~Tanabashi {\it et al.} [Particle Data Group],
  Review of Particle Physics.
  Phys.\ Rev.\ D {\bf 98}, 030001 (2018)

\bibitem{Niu:2018otv}
  J.~J.~Niu, L.~Guo and S.~M.~Wang,
  $HZ$ associated production with decay in the Alternative Left-Right Model at CEPC and future linear colliders
  Chin.\ Phys.\ C {\bf 42}, 093107 (2018)

\bibitem{Kiselev:1999sc}
  V.~V.~Kiselev, A.~K.~Likhoded and A.~I.~Onishchenko,
  Semileptonic $B_c$ meson decays in sum rules of QCD and NRQCD.
  Nucl.\ Phys.\ B {\bf 569}, 473 (2000)

\bibitem{Bodwin:1996tg}
  G.~T.~Bodwin, D.~K.~Sinclair and S.~Kim,
  Quarkonium decay matrix elements from quenched lattice QCD
  Phys.\ Rev.\ Lett.\  {\bf 77}, 2376 (1996)

\bibitem{Bagan:1994dy}
  E.~Bagan, H.~G.~Dosch, P.~Gosdzinsky, S.~Narison and J.~M.~Richard,
  Hadrons with charm and beauty.
  Z.\ Phys.\ C {\bf 64}, 57 (1994)

\bibitem{Bodwin:2002hg}
  G.~T.~Bodwin and A.~Petrelli,
  Order-$v^4$ corrections to $S$-wave quarkonium decay.
  Phys.\ Rev.\ D {\bf 66}, 094011 (2002)

\bibitem{Hahn:2000kx}
  T.~Hahn,
  Generating Feynman diagrams and amplitudes with FeynArts 3.
  Comput.\ Phys.\ Commun.\  {\bf 140}, 418 (2001)

\bibitem{Hahn:1998yk}
  T.~Hahn and M.~Perez-Victoria,
  Automatized one loop calculations in four-dimensions and D-dimensions.
  Comput.\ Phys.\ Commun.\  {\bf 118}, 153 (1999)

\bibitem{Heinemeyer:2013tqa}
  S.~Heinemeyer {\it et al.} [LHC Higgs Cross Section Working Group],
  Handbook of LHC Higgs Cross Sections: 3. Higgs Properties (2013).
  arXiv:1307.1347 [hep-ph]

\bibitem{Khachatryan:2016ctc}
  V.~Khachatryan {\it et al.} [CMS Collaboration],
  Search for Higgs boson off-shell production in proton-proton collisions at 7 and 8 TeV and derivation of constraints on its total decay width.
  JHEP {\bf 1609}, 051 (2016)

\bibitem{Bediaga:2012py}
  R.~Aaij {\it et al.} [LHCb Collaboration],
  Implications of LHCb measurements and future prospects.
  Eur.\ Phys.\ J.\ C {\bf 73}, 2373 (2013)

\bibitem{Bediaga:2018lhg}
  I.~Bediaga {\it et al.} [LHCb Collaboration],
  Physics case for an LHCb Upgrade II - Opportunities in flavour physics, and beyond, in the HL-LHC era (2018).
  arXiv:1808.08865 [hep-ex]


\bibitem{Brodsky:2011ta}
  S.~J.~Brodsky and X.~G.~Wu,
  Scale Setting Using the Extended Renormalization Group and the Principle of Maximum Conformality: the QCD Coupling Constant at Four Loops.
  Phys.\ Rev.\ D {\bf 85}, 034038 (2012)

\bibitem{Brodsky:2012rj}
  S.~J.~Brodsky and X.~G.~Wu,
  Eliminating the Renormalization Scale Ambiguity for Top-Pair Production Using the Principle of Maximum Conformality.
  Phys.\ Rev.\ Lett.\  {\bf 109}, 042002 (2012)

\bibitem{Mojaza:2012mf}
  M.~Mojaza, S.~J.~Brodsky and X.~G.~Wu,
  Systematic All-Orders Method to Eliminate Renormalization-Scale and Scheme Ambiguities in Perturbative QCD.
  Phys.\ Rev.\ Lett.\  {\bf 110}, 192001 (2013)

\bibitem{Brodsky:2013vpa}
  S.~J.~Brodsky, M.~Mojaza and X.~G.~Wu,
  Systematic Scale-Setting to All Orders: The Principle of Maximum Conformality and Commensurate Scale Relations.
  Phys.\ Rev.\ D {\bf 89}, 014027 (2014)


\end{thebibliography}
\bibliographystyle{unsrt}

\end{document}